\documentclass[aps,twocolumn,prb,showpacs]{revtex4-1}
\usepackage{graphicx}
\usepackage{amsmath,amssymb,amsfonts}
\usepackage[hyperindex,breaklinks]{hyperref}
\usepackage{bm}

\newcommand{\Ham}{\mathcal{H}}
\newcommand{\bea}{\begin{eqnarray}}
\newcommand{\eea}{\end{eqnarray}}
\newcommand{\beq}{\begin{equation}}
\newcommand{\eeq}{\end{equation}}
\newcommand{\nn}{\nonumber}

\begin{document}

\title{$B_{1g}$-like pairing states in two-leg ladder iron superconductors}
\author{Weicheng Lv}
\affiliation{Department of Physics and Astronomy, University of Tennessee, Knoxville, Tennessee 37996, USA}
\affiliation{Materials Science and Technology Division, Oak Ridge National Laboratory, Oak Ridge, Tennessee 37831, USA}
\author{Adriana Moreo}
\affiliation{Department of Physics and Astronomy, University of Tennessee, Knoxville, Tennessee 37996, USA}
\affiliation{Materials Science and Technology Division, Oak Ridge National Laboratory, Oak Ridge, Tennessee 37831, USA}
\author{Elbio Dagotto}
\affiliation{Department of Physics and Astronomy, University of Tennessee, Knoxville, Tennessee 37996, USA}
\affiliation{Materials Science and Technology Division, Oak Ridge National Laboratory, Oak Ridge, Tennessee 37831, USA}

\date{\today}

\begin{abstract}
Motivated by the recent report of superconductivity in Fe-based ladder materials, we study the pairing state
of a multi-orbital $t$-$J$ model defined on two-leg ladders using the standard mean-field theory. We find that
the superconducting order parameters change sign between the $d_{xz}$ and $d_{yz}$ orbitals in most of the phase
diagram. By analogy with the two-dimensional Fe planes, we conclude that the leading pairing channel of this state
belongs to the $B_{1g}$ symmetry class, which is
distinct from the common $s_{\pm}$ gap with the $A_{1g}$ symmetry.
By smoothly interpolating from planes into ladders, we show that a first-order transition occurs between
these two competing phases when the dimension of the system is reduced.
\end{abstract}


\maketitle


\section{Introduction}

Understanding the interplay between superconductivity and magnetism is the central topic
in the study of unconventional superconductors.\cite{Scalapino2012} In the context of the Fe-based materials
several theories have been proposed based on the weak-coupling
\cite{Mazin2008,Kuroki2008,Graser2009,Wang2009,Chubukov2008,Cvetkovic2009}
and strong-coupling\cite{Si2008,Seo2008,Chen2009a,Goswami2010,Nicholson2011} limits. In the former case,
near the spin-density-wave instability the pairing interaction is mediated by the exchange of
spin fluctuations between the hole and electron pockets. By contrast, theories that start in the strong-coupling
limit argue that superconductivity arises as in doped Mott insulators, where the electron pairing is induced
by the short-range magnetic exchanges between localized moments.
These two scenarios predict the same $s_{\pm}$ pairing state and are difficult to distinguish
because the electrons in the Fe-based superconductors
show both local and itinerant characters.\cite{Kou2009,Dai2009,Lv2010,Yin2010,Dai2012,Yu2013}
To resolve this controversy, one alternative approach is to explore new materials that may share a
universal pairing mechanism. Indeed, many significant insights have been gained from the
recent studies of $\rm{KFe_2As_2}$ that only has Fermi surface hole pockets\cite{Okazaki2012,Thomale2011a,Suzuki2011,Maiti2012a}
and $\rm{KFe_2Se_2}$ where only the electron pockets are present.\cite{Xu2012,Fang2011b,Yu2011b,Maier2011,Wang2011a,Dagotto2013}

In this regard, the very recent report\cite{Li2012a} of possible superconductivity in
the Fe-based two-leg ladder material $\rm{K_3Fe_4Se_6}$ may provide a new testing ground for the
existing theories. The actual sample of this material is made of a single layer of the bulk $\rm{KFe_2Se_2}$
grown along the crystal [110] direction. The basic building block is the $\rm{Fe_2Se_3}$ ladder,
which consists of two parallel Fe chains that are strongly bound together, as shown in Fig.~\ref{fig:ladder}.
These ladders are then weakly coupled to each other through layers of K atoms. In fact, the same Fe-based two-leg
ladder structure has been realized previously in bulk materials $\rm{BaFe_2Se_3}$ and $\rm{KFe_2Se_3}$, both of
which are, however, insulators. It is interesting to note that in $\rm{BaFe_2Se_3}$ and
below $T_N \approx 256$ K, the Fe spins form $2\times2$ ferromagnetic (FM) blocks, which are antiferromagnetically (AFM)
coupled along the chain direction.\cite{Caron2011,Lei2011,Nambu2012} For $\rm{KFe_2Se_3}$, below $T_N \approx 250$ K
there also exists a different  magnetic ground state\cite{Caron2012} in which the spins are FM aligned on the same rung,
but AFM ordered along the chain. Furthermore, both materials exhibit very large ordered moments,
ranging from 2.8 $\mu_\mathrm{B}$ per Fe in $\rm{BaFe_2Se_3}$ to 2.1 $\mu_\mathrm{B}$ in $\rm{KFe_2Se_3}$.
Assuming a valence +1 for K and -2 for Se, we find that superconducting $\rm{K_3Fe_4Se_6}$ has the Fe valence of +2.25,
which lies between +2 of $\rm{BaFe_2Se_3}$ and +2.5 of $\rm{KFe_2Se_3}$. Therefore superconductivity
in these Fe-ladder materials appears to arise from doping AFM insulators, which locates them closer
to the high-$T_c$ copper-oxide superconductors
than the earlier Fe-based pnictide superconductors.

In the context of the cuprates, extensive research has been carried out for the one-band model
on two-leg ladders.\cite{Dagotto1992,Sigrist1994,Dagotto1996,Dagotto1999}  It has been well established
that in the undoped limit there is a spin gap due to the particular ladder geometry, and that upon doping
superconductivity mainly occurs in the $d$-wave channel.\footnote{Note that $s$-wave and $d$-wave are not
the appropriate notations in the ladder system. But we will use them throughout the paper by analogy with
the two-dimensional system.} Both features are also found in the layered cuprates that are composed of
two-dimensional CuO$_2$ planes. So it is very interesting to study whether such a remarkable similarity
between the two-leg ladders and the two-dimensional planes for the case of the cuprates
also exists in the multi-orbital Fe-based superconductors.

The magnetic properties of the Fe ladders have been investigated theoretically using both
first principles\cite{Li2012b} and model calculations.\cite{Luo2013} In this work,
the pairing state of these materials will be studied. Because the parent compounds are insulators with
high N\'eel temperature and large ordered moment, in this publication we will follow the strong-coupling approach,
in particular, the mean-field theory of a multi-orbital $t$-$J$ model on two-leg ladders. We emphasize that,
by definition of the strong-coupling theory, the $t$-$J$ model should be derived from some Hubbard model
in the limit of large on-site interaction $U$. However, such procedure is very complicated due
to the multi-orbital nature of the Fe-based superconductors. Therefore, our current study and
other earlier works\cite{Seo2008,Goswami2010,Fang2011b,Yu2011b} simply start from a phenomenological
$t$-$J$ model by assuming the existence of the local moments. Our main result is that the
superconducting phase of the Fe ladders belongs to a different symmetry class from that of the two-dimensional Fe planes.
When dimension is reduced from planes to ladders, the system undergoes a first-order transition between the two phases.
The paper is organized as follows. First, the basic formulations are outlined in Sec.~\ref{sec:model}. The main results are
shown in Sec.~\ref{sec:results}. In particular, the isolated Fe ladder is considered in Sec.~\ref{sec:isolated} whereas
Sec.~\ref{sec:coupled} studies the nontrivial evolution from planes to ladders using a model system of coupled ladders.
Finally Sec.~\ref{sec:conclusion} concludes the publication with a discussion of results and a brief summary.

\section{Multi-orbital $t$-$J$ model on ladders}
\label{sec:model}

\begin{figure}
  \centering
  \includegraphics[width=8cm]{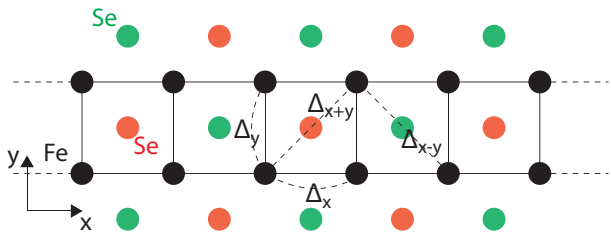}
  \caption{(Color online) Schematic illustration of the two-leg ladder iron selenides. The Se atoms are
located either above (red) or below (green) the Fe plane. $x$ and $y$ label the directions
of the chain and rung, respectively. $\Delta_x$, $\Delta_y$, $\Delta_{x+y}$, $\Delta_{x-y}$
are the superconducting order parameters with the pair of electrons located in nearest neighbors (NN)
sites along the chain or rung,
or involving next nearest neighbors (NNN) sites, respectively.}
  \label{fig:ladder}
\end{figure}

Let us first define the multi-orbital $t$-$J$ model on a two-leg ladder,
\beq
    \Ham = \Ham_K + \Ham_J.
\eeq

$\Ham_K$ is the kinetic energy, represented by a tight-binding model,
\beq
    \Ham_K = \sum_{\langle\bm{r},\bm{r}^\prime \rangle} \sum_{\alpha\beta, s} \left( t_{\bm{r}\bm{r}^\prime}^{\alpha\beta} c_{\bm{r}\alpha s}^\dagger c_{\bm{r}^\prime\beta s} + h.c. \right) - \mu \sum_{\bm{r}, \alpha} n_{\bm{r}}^\alpha,
\label{eq:HK}
\eeq
where $c_{\bm{r}\alpha s}^\dagger$ creates an electron at site $\bm{r}$ on orbital $\alpha$ with spin $s$.
The density operator is $n_{\bm{r}}^\alpha = \sum_s c_{\bm{r}\alpha s}^\dagger c_{\bm{r}\alpha s}$, and $\mu$ is the chemical potential. Because our model is defined on a two-leg ladder as illustrated in Fig.~\ref{fig:ladder}, the site index $\bm{r}$ has two components, $\bm{r} = (m,i)$, where $m=1,2$ labels the two parallel chains and $i$ is the site index on each chain. It should also be noted that the whole system does not have the full translational symmetry of the Fe ladder along the $x$ (chain) direction due to the staggered positions of the Se atoms (Fig.~\ref{fig:ladder}). The real unit cell of the $\rm{Fe_2Se_3}$ ladder contains four Fe sites. However, similarly as carried out in higher-dimensional systems,\cite{Lee2008,Calderon2009,Daghofer2010,Lv2011a} we can perform a gauge transformation, $c_{\bm{r}\alpha s} \rightarrow \exp\left( i \bm{K}\cdot\bm{r} \right)c_{\bm{r}\alpha s}$ for $\alpha = d_{xy}, d_{x^2-y^2}, d_{3z^2-r^2}$, where $\bm{K} = (\pi,\pi)$. After such a transformation, the tight-binding model $\Ham_K$ (\ref{eq:HK}) becomes invariant under the translation by one Fe lattice spacing along the $x$ (chain) direction. Therefore, we can work in the effective unit cell that contains only two Fe sites along the rung. The hopping parameters $t_{\bm{r}\bm{r}^\prime}^{\alpha\beta}$ are taken from the earlier three-orbital model\cite{Daghofer2010} with $d_{xz}$, $d_{yz}$, and $d_{xy}$ orbitals. In principle, due to the ladder geometry, these parameters can have different values here.\cite{Luo2013} Especially, the hopping amplitudes along the $x$ (chain) direction are not necessarily the same as those along the $y$ (rung) direction. The on-site energies of the $d_{xz}$ and $d_{yz}$ orbitals should also be different.
However, for the purpose of simplicity, we use the same set of parameters derived for the two-dimensional Fe planes
in Ref.~\onlinecite{Daghofer2010}. Namely, the tight-binding parameters themselves retain the $C_4$ rotational
symmetry of the square lattice, and it is the geometry of the ladder that induces the symmetry breaking.
This procedure will facilitate the discussion on the evolution of the superconducting state from planes to ladders.
We also note that the band structure of the Fe ladder in our model, as shown in Fig.~\ref{fig:band},
is different from that obtained by the first-principles calculation in Ref.~\onlinecite{Li2012a}. However,
while our main conclusion, i.e. the appearance
of the $B_{1g}$ pairing state in ladders, may depend on the particular hopping parameters chosen here,
our main goal is to show that at least in \emph{one} case a change in pairing symmetry can occur
in the interpolation from planes to ladders. Developing a
more refined set of parameters from first-principles calculations would not be appropriate considering that
the many-body techniques employed here are based on the strong-coupling limit.
Finally we set the chemical potential $\mu$ to give an electron filling $n=3.75$ per site, which corresponds to 0.25 hole doping
as in the superconducting Fe-ladder material $\rm{K_3Fe_4Se_6}$.

$\Ham_J$ is the exchange energy between the spins on the neighboring sites,
\beq
    \Ham_J  =  \sum_{\langle\bm{r},\bm{r}^\prime \rangle} \sum_{\alpha\beta} J_{\bm{r}\bm{r}^\prime}^{\alpha\beta} \left( \bm{S}_{\bm{r}}^\alpha \cdot \bm{S}_{\bm{r}^\prime}^\beta - \frac{1}{4}
    n_{\bm{r}}^\alpha n_{\bm{r}^\prime}^\beta \right),
\label{eq:HJ}
\eeq
where the spin operator is $\bm{S}_{\bm{r}}^\alpha = 1/2 \sum_{s s^\prime} c_{\bm{r}\alpha s}^\dagger \bm{\sigma}_{s s^\prime} c_{\bm{r}\alpha s^\prime}$, with $\bm{\sigma}$ being the Pauli matrix.
Similarly as in earlier work,\cite{Si2008} the superexchanges are restricted to involve only
the nearest-neighbors (NN), $J_1$, and the next-nearest-neighbors (NNN), $J_2$. It is further assumed
that the NN exchanges are isotropic between the bonds along the chain and those along the rung,
$J_{1x} = J_{1y} = J_1$, which does not necessarily hold for the ladder geometry but our intention is
to reduce to the minimum the number of free parameters in the model.

Following the standard procedure,\cite{Seo2008,Goswami2010,Fang2011b,Yu2011b}
we define the singlet pairing operators between the sites $\bm{r}$ and $\bm{r}^\prime$,
\beq
    P_{\bm{r}\bm{r}^\prime}^{\alpha\beta} = c_{\bm{r}\alpha\uparrow} c_{\bm{r}^\prime\beta\downarrow} - c_{\bm{r}\alpha\downarrow} c_{\bm{r}^\prime\beta\uparrow},
\eeq
and express the exchange energy $\Ham_J$ in terms of $P_{\bm{r}\bm{r}^\prime}^{\alpha\beta}$,
\beq
    \Ham_J =  - \sum_{\langle\bm{r},\bm{r}^\prime \rangle} \sum_{\alpha\beta}  \frac{J_{\bm{r}\bm{r}^\prime}^{\alpha\beta}}{2} \left( P_{\bm{r}\bm{r}^\prime}^{\alpha\beta} \right)^\dagger P_{\bm{r}\bm{r}^\prime}^{\alpha\beta}.
\eeq
Then, $\Ham_J$ can be decoupled by the standard mean-field theory in the pairing channel.
The superconducting order parameters between orbitals $\alpha$ and $\beta$ at sites $\bm{r}$ and $\bm{r}^\prime$ are defined as
\beq
    \Delta_{\bm{r}\bm{r}^\prime}^{\alpha\beta} = J_{\bm{r}\bm{r}^\prime}^{\alpha\beta} \left\langle P_{\bm{r}\bm{r}^\prime}^{\alpha\beta} \right\rangle.
\eeq
For the $J_1$-$J_2$ model used here, assuming translational symmetry, we can identify four distinct order
parameters for each orbital pair ($\alpha$, $\beta$), namely $\Delta_x$ and $\Delta_y$ between
NN sites along the chain and rung, and $\Delta_{x+y}$ and $\Delta_{x-y}$ between NNN sites, as shown in
Fig.~\ref{fig:ladder}. After the standard Fourier transform, we have the following BCS mean-field Hamiltonian,
\bea
    \Ham_{MF} & = &
    \sum_{k,mn,\alpha\beta,s} \left( \varepsilon_{mn}^{\alpha\beta}(k) - \mu\delta_{\alpha\beta}\delta_{mn} \right)c_{km\alpha s}^\dagger c_{kn\beta s}  \nn \\
    & & \, + \sum_{k,mn,\alpha\beta} \left( \Delta_{mn}^{\alpha\beta}(k) c_{km\alpha\uparrow}^\dagger c_{-kn\beta\downarrow}^\dagger + h.c. \right).
\label{eq:HBCS}
\eea
Such a model can be solved self-consistently. In general, with different initial conditions, there are many
self-consistent solutions, of which we choose the one that minimizes the ground state energy,\cite{Goswami2010,Fang2011b,Yu2011b}
\beq
     \mathcal{E} = \sum_{\langle\bm{r},\bm{r}^\prime \rangle} \sum_{\alpha\beta} \frac{1}{2J_{\bm{r}\bm{r}^\prime}^{\alpha\beta}}  \left\vert \Delta_{\bm{r}\bm{r}^\prime}^{\alpha\beta} \right\vert^2 - \sum_{k,n} \left( E_{k,n} - \varepsilon_{k,n} + \mu \right).
\eeq
$E_{k,n}$ are the quasiparticle energies
obtained from the diagonalization of the BCS mean-field Hamiltonian $\Ham_{MF}$ (\ref{eq:HBCS}),
whereas $\varepsilon_{k,n}$ are the band dispersions of the tight-binding model $\Ham_{K}$ (\ref{eq:HK}).
In order to limit the number of parameters, we assume that the superexchanges $J$ are orbitally diagonal
and orbitally independent, namely $J_{\bm{r}\bm{r}^\prime}^{\alpha\beta} = \delta_{\alpha\beta} J_{\bm{r}\bm{r}^\prime}$.
Therefore the pairing only occurs within the same orbitals,
$\Delta_{\bm{r}\bm{r}^\prime}^{\alpha\beta} = \delta_{\alpha\beta} \Delta_{\bm{r}\bm{r}^\prime} (\alpha)$.
The case in which the intra-orbital and inter-orbital superexchanges are of equal
strength has also been studied. Similar to earlier works,\cite{Goswami2010,Fang2011b} the inclusion of the inter-orbital
pairings, which are found to be much smaller than the intra-orbital ones, does not change the results
qualitatively at least within the approximations used here. Thus, for simplicity,
we will only focus on the intra-orbital pairing arising from orbitally
diagonal superexchange interactions in this article.

\section{Results}
\label{sec:results}

\subsection{Isolated ladders}
\label{sec:isolated}

\begin{figure}
  \centering
  \includegraphics[width=8cm]{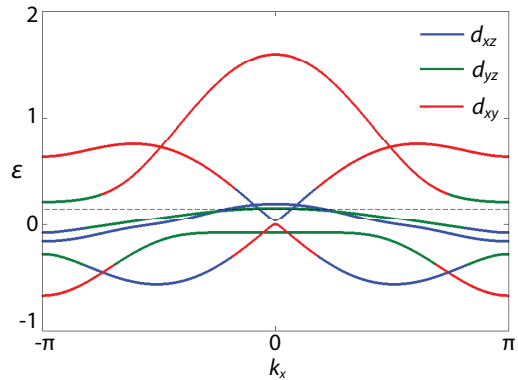}
  \caption{(Color online) Band dispersions along the $x$ (chain) direction. The colors correspond to the dominant
orbital contributions, $d_{xz}$ (blue), $d_{yz}$ (green), and $d_{xy}$ (red). The horizontal dashed line denotes
the chemical potential for filling $n=3.75$.}
  \label{fig:band}
\end{figure}

Let us consider first a single Fe ladder as shown in Fig.~\ref{fig:ladder}.
Periodic and open boundary conditions are imposed along the chain and rung directions, respectively.
The band dispersions $\varepsilon_{k,n}$ along the $x$ (chain) direction are plotted in Fig.~\ref{fig:band}
with the colors denoting the dominant orbital contributions. From now on, the implicit energy unit of $e$V will be
used unless noted otherwise. Since the effective unit cell contains two Fe sites along the $y$ (rung) direction,
there are six bands in the present three-orbital model. As discussed earlier, we have set the chemical potential
at the filling level $n=3.75$, corresponding to 0.25 hole doping. Apparently, no nesting instability
can be identified, with the $d_{xz}$ and $d_{yz}$ orbitals dominating the states at the Fermi energy.
Because the ladder structure explicitly breaks the symmetry of the two-dimensional square lattice, one
cannot rigorously categorize the pairing state by the one-dimensional representation of the $D_{4h}$ point
group, as in earlier efforts.\cite{Goswami2010,Fang2011b,Yu2011b,Daghofer2010}
For ladders it is necessary to study each of the superconducting order parameters
$\Delta_{\bm{r}\bm{r}^\prime}(\alpha)$ separately. However, in order to compare the ladders
with the planes, we will attempt to identify approximately  the leading pairing symmetry
by analogy with the two-dimensional case.

\begin{figure}
  \centering
  \includegraphics[width=8cm]{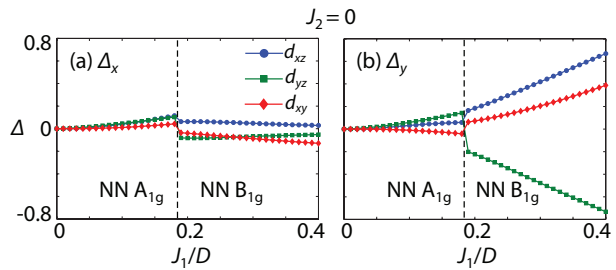}
  \caption{(Color online) Superconducting order parameters (a) $\Delta_x$ and (b) $\Delta_y$ for each orbital as functions
of the NN exchange $J_1$ for the case of the NNN exchange $J_2=0$. $D$ is the bandwidth. The vertical dashed line marks the
first-order transition between the two phases, i.e.~NN $A_{1g}$ and NN $B_{1g}$.}
  \label{fig:Delta1}
\end{figure}

As the first step, let us set the NNN exchange $J_2$ to zero and consider only the NN pairing induced by $J_1$.
In Fig.~\ref{fig:Delta1}, the superconducting order parameters between NN sites along the $x$ (chain)
and $y$ (rung) directions, $\Delta_x$ and $\Delta_y$, are shown for each of the three orbitals as functions of $J_1$.
The coupling $J_1$ is measured in units of the bandwidth $D$. Two different regimes can be identified
as $J_1$ is varied. When $J_1 \lesssim 0.18D$, $\Delta_x$ and $\Delta_y$ have comparable amplitudes.
In particular, they take the same sign on the $d_{xz}$ and $d_{yz}$ orbitals ($s$-wave) whereas they
have different signs ($d$-wave) on $d_{xy}$. At first sight, it is difficult to assign the pairing symmetry
for such a state because different symmetries occur on different orbitals. However, we note that the electrons
in the vicinity of the Fermi energy are mostly from the $d_{xz}$ and $d_{yz}$ orbitals (Fig.~\ref{fig:band}).
Consequently the superconducting gaps on these two orbitals are larger than the gap on $d_{xy}$ so that the system
can gain more condensation energy. Therefore, we can approximately categorize the pairing symmetry as $A_{1g}$
according to the gap structure restricted to the $d_{xz}$ and $d_{yz}$ orbitals. However, note that when $J_1 \gtrsim 0.18D$, the system
evolves into a second phase through a first-order transition. The superconducting gap remains $s$-wave
on $d_{xz}$ and $d_{yz}$, and $d$-wave on $d_{xy}$. But unlike the previous case, $\Delta(d_{xz})$ and $\Delta(d_{yz})$
now display opposite signs. We recall that the overall pairing symmetry of a multi-orbital system depends on both the spatial and orbital gap structures.\cite{Goswami2010,Fang2011b,Yu2011b,Daghofer2010} Therefore, by analogy with the two-dimensional planes, this phase belongs to the $B_{1g}$ symmetry
class due to the sign difference between $\Delta(d_{xz})$ and $\Delta(d_{yz})$. As $J_1$ further increases, $\Delta_x$ gets strongly suppressed while $\Delta_y$
becomes the leading pairing channel. This result can be understood from the earlier work on the one-band model in
the context of the cuprates.\cite{Dagotto1992,Sigrist1994,Dagotto1996,Dagotto1999} In the large $J_1$ limit,
the undoped system has a spin-liquid ground state in which the spins tend to form local singlets on the same rung.
When holes are doped into the system, they prefer to occupy the sites on the same rung because such a state has the
lowest energy by breaking the fewest local singlets. Therefore, the pairing mainly occurs in the $\Delta_y$ channel
along the rung when $J_1$ is large. Here we have found similar results using a multi-orbital model.

\begin{figure}
  \centering
  \includegraphics[width=8cm]{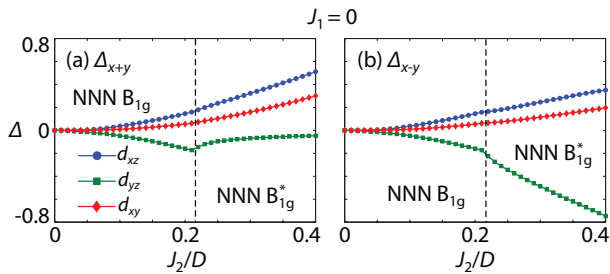}
  \caption{(Color online) Superconducting order parameters (a) $\Delta_{x+y}$ and (b) $\Delta_{x-y}$ for
each orbital as functions of the NNN exchange $J_2$ for the case of the NN exchange $J_1=0$. $D$ is the bandwidth.
The vertical dashed line marks the second-order transition between the two phases, i.e.~NNN $B_{1g}$ ($\Delta_{x+y} = \Delta_{x-y}$)
and NNN $B_{1g}^*$ ($\Delta_{x+y} \neq \Delta_{x-y}$).}
  \label{fig:Delta2}
\end{figure}

Now let us consider the other limit, namely the case where the NN exchange is zero, $J_1= 0$, and the pairing occurs
between the NNN sites. Figure \ref{fig:Delta2} shows the NNN superconducting order parameters
$\Delta_{x+y}$ and $\Delta_{x-y}$ as functions of $J_2$. In this case there are two different phases
stabilized at the small and large $J_2$ limits, respectively. When $J_2 \lesssim 0.21D$, the two NNN gaps
are equal ($s$-wave) for each of the three orbitals. Similarly to the previous case, we identify the dominant
pairing symmetry by the gap structure on the leading $d_{xz}$ and $d_{yz}$ orbitals. Because $\Delta_{x \pm y}$ have
opposite signs on $d_{xz}$ and $d_{yz}$, the $B_{1g}$ symmetry is assigned to this phase.
When $J_2 \gtrsim 0.21D$, $\Delta_{x+y}$ and $\Delta_{x-y}$ now have different values, which suggests
a mixture phase with both $s$- and $d$-wave components. Similar results with $\Delta_{x+y} \neq \Delta_{x-y}$
have also been found in the study of the two-dimensional Fe planes in the large $J_2$ limit.\cite{Fang2011b,Yu2011b}
It is important to emphasize that $B_{1g}$ is still the dominant pairing symmetry here although other gap amplitudes with different
symmetries also take significant values. Consequently, this phase is labeled as $B_{1g}^*$ to distinguish it
from the previous $B_{1g}$ phase at $J_2 \lesssim 0.21D$. In contrast to the first-order transition from the
NN $A_{1g}$ to NN $B_{1g}$ states observed in Fig.~\ref{fig:Delta1}, here the leading pairing symmetry does not change
between the NNN $B_{1g}$ and NNN $B_{1g}^*$ phases, and thus the transition occurs smoothly,
as shown in Fig.~\ref{fig:Delta2}. Finally, we recall that in the earlier
works\cite{Seo2008,Goswami2010,Fang2011b,Yu2011b} for the two-dimensional Fe planes, the NNN $J_2$ coupling consistently
led to the $s_{\pm}$ gap with the $A_{1g}$ symmetry, in which the NNN superconducting order parameters
are equal not only on the two NNN bonds, but also on the two orbitals of $d_{xz}$ and $d_{yz}$, i.e.~
$\Delta_{x \pm y} (d_{xz}) = \Delta_{x \pm y} (d_{yz})$. Consequently, in our calculations the pairing symmetry of the planes
is fundamentally different from that of the two-leg ladders, where a sign change occurs between the superconducting
order parameters on $d_{xz}$ and $d_{yz}$. This nontrivial dimensional crossover from planes to ladders
will be studied in detail in Sec.~\ref{sec:coupled}.

\begin{figure}
  \centering
  \includegraphics[width=8cm]{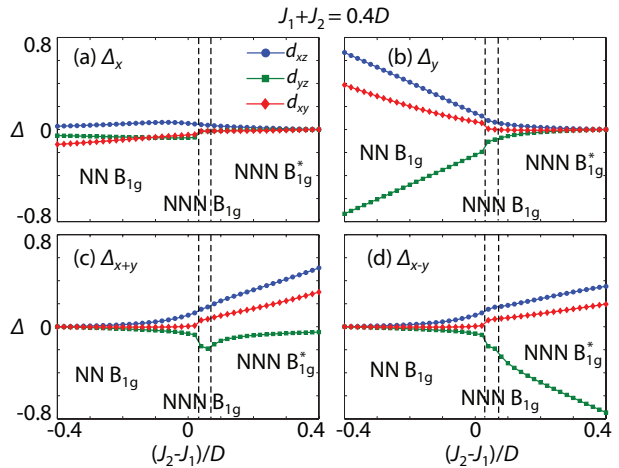}
  \caption{(Color online) Superconducting order parameters (a) $\Delta_{x}$, (b) $\Delta_{y}$, (c) $\Delta_{x+y}$, and (d) $\Delta_{x-y}$ for each orbital as functions of the difference between the NNN and NN exchanges, $J_2-J_1$.
The sum of $J_1$ and $J_2$ is fixed to be $0.4D$, where $D$ is the bandwidth. The vertical dashed lines separate the three phases, NN $B_{1g}$, NNN $B_{1g}$, and NNN $B_{1g}^*$.}
  \label{fig:Delta12}
\end{figure}

Thus far, we have observed that two different phases, the NN $B_{1g}$ state dominated by the rung
pairing $\Delta_y$ and the NNN $B_{1g}^*$ state with $\Delta_{x+y} \neq \Delta_{x-y}$, are stabilized
at the limits of large $J_1$ and large $J_2$, respectively. It will be interesting to investigate
how the two phases compete with each other when both $J_1$ and $J_2$ are nonzero. For this purpose,
let us fix the sum of $J_1$ and $J_2$ to be $J_1+J_2 = 0.4D$, and plot each of the four superconducting
order parameters as a function of $J_2-J_1$ in Fig.~\ref{fig:Delta12}. In this case, we observe
that the NN rung pairing $\Delta_y$ is still the dominant pairing channel even when a small but
finite $J_2$ is turned on. This NN $B_{1g}$ phase extends up to the parameter space with $J_2 - J_1 \lesssim 0.03D$.
Then, the system transitions into the NNN $B_{1g}$ state with $\Delta_{x+y} = \Delta_{x-y}$
when $0.03D \lesssim J_2 - J_1 \lesssim 0.07D$, before the $B_{1g}^*$ phase ($\Delta_{x+y} \neq \Delta_{x-y}$)
becomes the ground state at $J_2 - J_1 \gtrsim 0.07D$. It is clear from Fig.~\ref{fig:Delta12} that
the transition between the NN $B_{1g}$ and the NNN $B_{1g}$ states is characterized by a discontinuous
jump of the order parameters. However, because the two phases belong to the same $B_{1g}$ symmetry class, the
transition is not always first order and a smooth crossover does occur at smaller superexchanges $J$ (results not shown here).

\begin{figure}
  \centering
  \includegraphics[width=8cm]{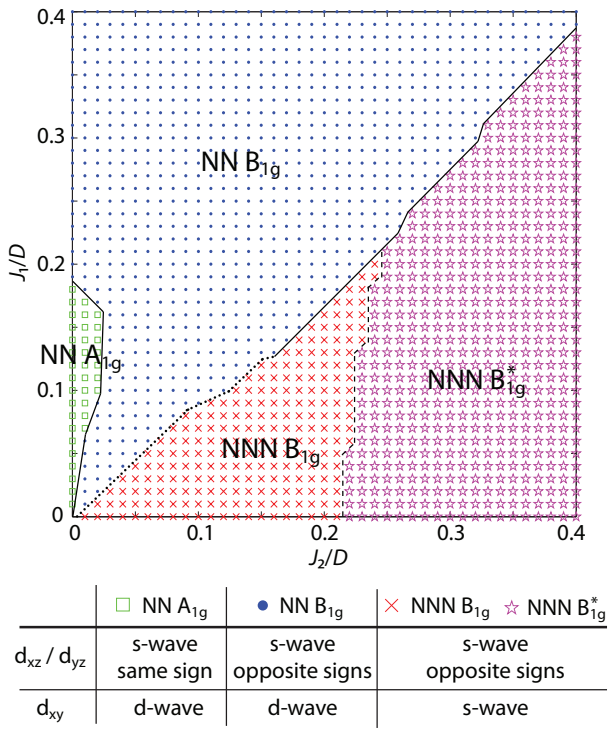}
  \caption{(Color online) The superconducting phase diagram of the isolated Fe ladders as a function of $J_1$ and $J_2$,
both measured in units of the bandwidth $D$. Different symbols correspond to different phases determined
by the leading pairing symmetries. The solid line marks a first-order transition; the dashed line
represents a second-order transition; and the dotted line illustrates the crossover between the NN $B_{1g}$ and the NNN $B_{1g}$ phases at small $J$.}
  \label{fig:phase}
\end{figure}

Finally, we have systematically studied the ground state of the system as a function of $J_1$ and $J_2$ by varying
each of those couplings from 0 to $0.4D$. The phase diagram, which is determined by the leading pairing
symmetries, is shown in Fig.~\ref{fig:phase}. We find that the NN $A_{1g}$ phase, which was observed at $J_1 \lesssim 0.18D$
with $J_2=0$ (Fig.~\ref{fig:Delta1}), only exists in a limited region with very small NNN exchange $J_2 \lesssim 0.03D$.
Instead, the phase diagram is dominated by the NN $B_{1g}$ and the NNN $B_{1g}$  ($B_{1g}^*$) phases,
which are separated approximately by the line $J_1 \approx J_2$. When $J \gtrsim 0.16D$ (solid line),
the transition between the two phases is first order as shown in Fig.~\ref{fig:Delta12}. The intermediate
NNN $B_{1g}$ phase disappears for $J\gtrsim 0.25D$, where a direct transition occurs between the NN $B_{1g}$
and NNN $B_{1g}^*$ phases. Furthermore, the NNN $B_{1g}$ and NNN $B_{1g}^*$ states are separated by a nearly
vertical dashed line around $J_2 \approx 0.23D$, where a second-order phase transition takes place with a
smooth onset of the pairing amplitudes with different symmetries. As discussed previously, the pairing
symmetry does not change from the NN $B_{1g}$ to the NNN $B_{1g}$ states. Consequently the transition
between the two phases is not necessarily first order.
Indeed, when $J \lesssim 0.16D$, we do observe a crossover transition (dotted line), where the superconducting order parameters exhibit
continuous changes. By contrast, a first-order
transition (solid line) always occur between the NN $A_{1g}$ and NN $B_{1g}$ phases, which belong to different symmetry classes.

\subsection{Coupled ladders}
\label{sec:coupled}

As the discussions in Sec.~\ref{sec:isolated} have shown, for the two-leg Fe ladders, $B_{1g}$ is the dominant
pairing symmetry and the superconducting order parameters have opposite signs on the $d_{xz}$ and $d_{yz}$ orbitals.
On the other hand, for the two-dimensional Fe planes,\cite{Seo2008,Goswami2010,Fang2011b,Yu2011b} the NNN exchange
$J_2$ consistently produces the $s_\pm$ gap with the $A_{1g}$ symmetry, which requires $\Delta_{x\pm y} (d_{xz}) = \Delta_{x\pm y} (d_{yz})$.
Apparently, the superconducting states of the Fe ladders and the Fe planes are fundamentally different.

\begin{figure}
  \centering
  \includegraphics[width=6cm]{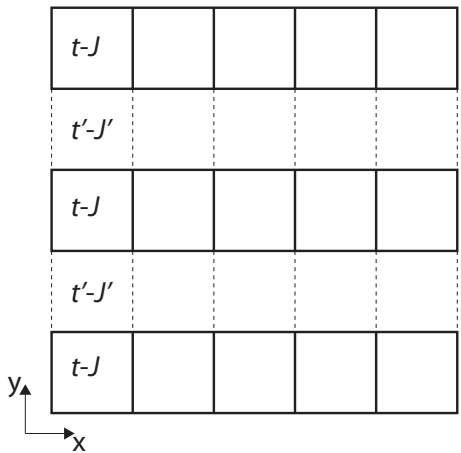}
  \caption{Illustration of the coupled ladders used here. The system consists of two-leg ladders aligned
along the $x$ direction (solid lines). These ladders are coupled along the $y$ direction through weaker
inter-ladder couplings (dashed lines). $t$ ($t^\prime$) and $J$ ($J^\prime$) are intra-ladder (inter-ladder)
hopping amplitudes and exchange constants, respectively.}
  \label{fig:coupled}
\end{figure}

In this section, we will study how this change of symmetry occurs as the system evolves from planes to ladders.
For this purpose, let us consider the model system shown in Fig.~\ref{fig:coupled}. It is a two-dimensional
square lattice consisting of $N \times N$ sites with periodic boundary conditions in both directions.
Translational symmetry is imposed along the $x$ direction, while strong and weak bonds occur in staggered
order along the $y$ direction. Then, essentially we have $N/2$ parallel two-leg ladders that are weakly
coupled to their neighbors. We use $t$ and $J$ for the intra-ladder hopping amplitudes and superexchanges,
and $t^\prime$ and $J^\prime$ for the inter-ladder ones. For simplicity, we have chosen $t^\prime = \eta t$
and $J^\prime = \eta J$, where the ratio $\eta$ is a single parameter representing the relative strength of
the inter-ladder couplings with respect to the intra-ladder ones. In the limit of $\eta = 1$, the system
becomes an isotropic two-dimensional square lattice, which has been extensively studied in
earlier works.\cite{Seo2008,Goswami2010,Fang2011b,Yu2011b} When $\eta = 0$, the system reduces to $N/2$ isolated ladders that were
considered previously in Sec.~\ref{sec:isolated}. By varying $\eta$ from 1 to 0, we can tune the system
continuously from two-dimensional planes to two-leg ladders, and study how superconductivity evolves.
As the band dispersion changes with $\eta$, the chemical potential $\mu$ needs to be adjusted accordingly so that
the filling level is always fixed at $n=3.75$. The exchange constants $J_1$ and $J_2$ are always measured in units
of $D$, the bandwidth of the isolated ladders.

\begin{figure}
  \centering
  \includegraphics[width=8cm]{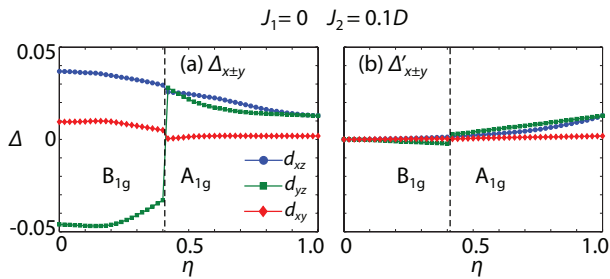}
  \caption{(Color online) The NNN intra-ladder and inter-ladder superconducting order parameters,
(a) $\Delta_{x\pm y}$ and (b) $\Delta^\prime_{x \pm y}$, for each orbital as functions of the relative
coupling strength parameter $\eta$. The exchange constants are fixed to values $J_1 =0$ and $J_2 =0.1D$.
 The vertical dashed line marks the first-order transition between the $A_{1g}$ phase of the planes and the $B_{1g}$ phase of the ladders.}
  \label{fig:etaNNN1}
\end{figure}

Let us first consider the role played by the NNN exchange $J_2$ by setting the NN exchange to $J_1=0$.
The superconducting order parameters at $J_2 =0.1D$ are plotted as functions of $\eta$ in Fig.~\ref{fig:etaNNN1},
with $\Delta$ and $\Delta^\prime$ representing the intra-ladder and inter-ladder pairing strengths, respectively.
For the parameters used here, we always have $\Delta_{x+y} = \Delta_{x-y}$ and $\Delta^\prime_{x+y} = \Delta^\prime_{x-y}$,
so only one of each is shown. As it can be observed from Fig.~\ref{fig:etaNNN1}, when $\eta=1$ the intra-ladder
and the inter-ladder pairing strengths are equal,
i.e.~$\Delta_{x\pm y} = \Delta^\prime_{x \pm y}$, which preserves
the translational symmetry along the $y$ direction. Furthermore, these superconducting order parameters
take the same value on the $d_{xz}$ and $d_{yz}$ orbitals. Such a state belongs to the $A_{1g}$ symmetry
class, with the pairing structural factor $\cos{k_x}\cos{k_y}$, in agreement with
earlier studies.\cite{Goswami2010,Fang2011b,Yu2011b}
When $\eta$ is reduced, the translational symmetry
along the $y$ direction is now explicitly broken, with the intra-ladder pairing being enhanced and the inter-ladder
one being reduced. But $\Delta_{x \pm y} (d_{xz})$ and $\Delta_{x \pm y} (d_{yz})$ still have the same sign
as long as $\eta \gtrsim 0.41$. Across $\eta \approx 0.41$, the system undergoes a discontinuous transition
from the $A_{1g}$ phase into the $B_{1g}$ phase, where the signs of $\Delta_{x \pm y} (d_{xz})$
and $\Delta_{x \pm y} (d_{yz})$ suddenly become opposite. Then the system evolves continuously
into the state of the isolated ladders that was  previously discussed at $\eta = 0$. Indeed the superconducting
phases of the two-dimensional planes and two-leg ladders belong to different symmetry classes. So the system
does not evolve smoothly from planes to ladders, but
a first-order transition separates the two competing phases.

\begin{figure}
  \centering
  \includegraphics[width=8cm]{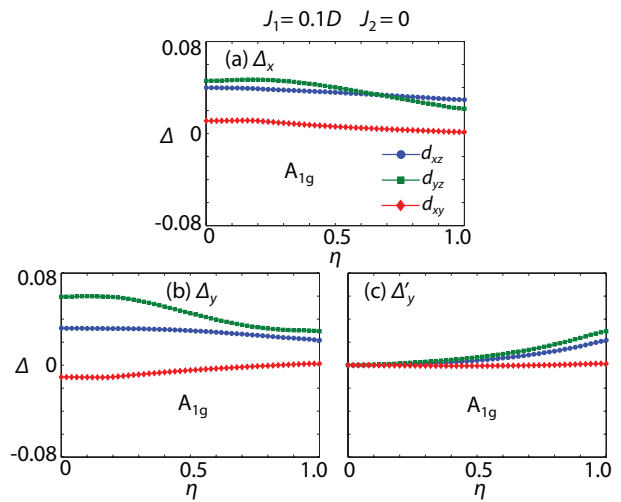}
  \caption{(Color online)The NN intra-ladder and inter-ladder superconducting order parameters,
(a) $\Delta_x$, (b) $\Delta_y$, and (c) $\Delta^\prime_y$ for each orbital as functions of the
relative coupling strength parameter $\eta$. The exchange constants are fixed to values
$J_1 =0.1D$ and $J_2 =0$. In this case the system transitions smoothly from planes to ladders,
where $A_{1g}$ is always the dominant pairing symmetry.}
  \label{fig:etaNN1}
\end{figure}

Now we focus on the NN pairings by setting $J_1=0.1D$ and $J_2 =0$. The results are shown
in Fig.~\ref{fig:etaNN1}, where the intra-ladder $\Delta_x$, $\Delta_y$ and the inter-ladder $\Delta^\prime_y$ are displayed.
Again at $\eta=1$, the intra-ladder and inter-ladder pairings are equal, $\Delta_y = \Delta_y^\prime$. Furthermore,
on the $d_{xz}$ and $d_{yz}$ orbitals, we have $\Delta_x(d_{xz}) = \Delta_y(d_{yz})$ and $\Delta_x(d_{yz}) = \Delta_y(d_{xz})$.
Then, this state also belongs to the $A_{1g}$ symmetry class with the leading structural
factor $\cos{k_x}+\cos{k_y}$. Similarly to the previous case, when $\eta$ is reduced the intra-ladder
pairings are enhanced while the inter-ladder ones are suppressed. However, we do not observe
any discontinuity as $\eta$ varies from 1 to 0, which suggests that the superconducting phase
of the planes and that of the ladders are smoothly connected with one another. Actually, as shown by Fig.~\ref{fig:Delta1},
the leading pairing channel of the two-leg ladders also has the $A_{1g}$ symmetry at $J_1=0.1D$.
Then, the evolution from planes to ladders is continuous without modifications of the pairing symmetry.
However, it should be noted that the superconducting order parameter on the $d_{xy}$ orbital does change
from $s$-wave in the planes to $d$-wave in the ladders. But this symmetry change on the $d_{xy}$ orbital
only occurs continuously because the pairing symmetry is mostly determined by the gap structures on the $d_{xz}$ and $d_{yz}$ orbitals, which dominate the states around the Fermi energy.

\begin{figure}
  \centering
  \includegraphics[width=8cm]{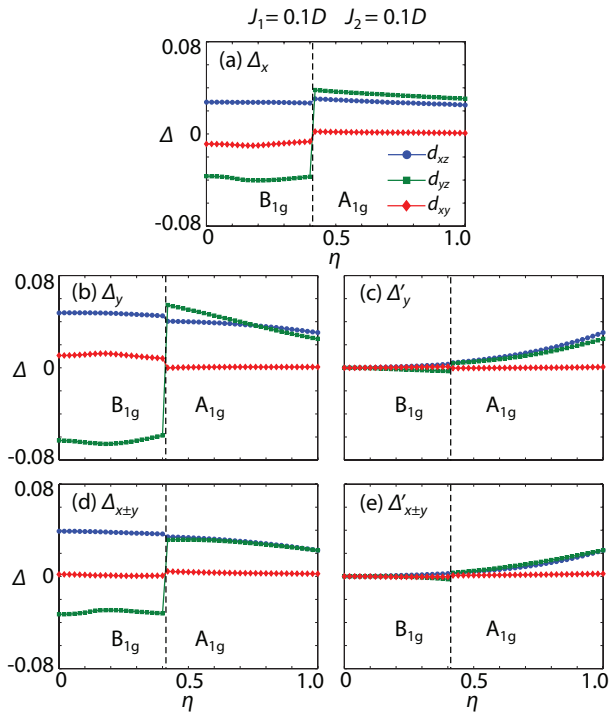}
  \caption{(Color online) The superconducting order parameters, (a) $\Delta_x$, (b) $\Delta_y$,
(c) $\Delta^\prime_y$, (d) $\Delta_{x\pm y}$, and (e) $\Delta^\prime_{x \pm y}$ for each orbital
as functions of the relative coupling strength parameter $\eta$. The exchange constants have been set
to $J_1 =0.1D$ and $J_2 =0.1D$. The vertical dashed line marks the first-order transition between the
$A_{1g}$ phase of the planes and the $B_{1g}$ phase of the ladders.}
  \label{fig:etaNN1NNN1}
\end{figure}

Finally, let us study the case in which both $J_1$ and $J_2$ are present. Figure \ref{fig:etaNN1NNN1}
shows all five independent superconducting order parameters as functions of $\eta$ at fixed $J_1 = J_2 =0.1D$.
According to the phase diagram in Fig.~\ref{fig:phase}, for the isolated ladders at $\eta=0$, the ground state
is a mixture of the NN $B_{1g}$ phase and the NNN $B_{1g}$ phase ($\Delta_{x+y} = \Delta_{x-y}$), both of which
are characterized by opposite signs of $\Delta(d_{xz})$ and $\Delta(d_{yz})$. By contrast, for the isotropic
planes at $\eta=1$, the system belongs to the $A_{1g}$ symmetry class with both the NN and NNN pairings.
As explained earlier, the evolution from planes to ladders cannot be smooth due to
the symmetry change as $\eta$ varies. Indeed our calculation results in Fig.~\ref{fig:etaNN1NNN1}
confirmed this prediction, with a sharp first-order transition occurring at $\eta \approx 0.41$.
Furthermore, as the phase diagram in Fig.~\ref{fig:phase} shows, in most portions of the $J_1$-$J_2$
phase diagram, $B_{1g}$ is the leading pairing symmetry of the two-leg Fe ladders. The NN $A_{1g}$ phase
only occupies a limited portion, where the values of $J_1$ and $J_2$ are likely outside the physically
relevant regime for the Fe-based superconductors. Then, it can be concluded that the evolution
from planes to ladders in the model for Fe-based superconductors considered here
occurs through a first-order transition with a
nontrivial modification of the pairing symmetry.

Before ending this subsection, it is important to point out that our analysis is restricted to
the case of small superexchanges $J$, where the ground state of the superconducting Fe planes
automatically preserves the translational symmetry along the $y$ direction. Namely, we always
have $\Delta = \Delta^\prime$ at $\eta=1$ although they are treated as independent variables.
However, when $J$ becomes sufficiently large, an inhomogeneous solution with $\Delta \neq \Delta^\prime$
was found to have a lower energy, spontaneously breaking the translation symmetry along the $y$ direction
at $\eta = 1$. We believe that this ``stripe''-like state is nevertheless an artifact of the two-site unit
cell used in our calculation. In fact, by enlarging the unit cell to a 2$\times$2 plaquette, we have observed
that the favorable solution becomes a checkerboard state, which breaks the translational symmetry along both
$x$ and $y$, but preserves the $C_4$ rotational symmetry. It is certainly possible that further inhomogeneous
solutions with lower energies can be found if we keep enlarging the unit cell. More importantly, the gap
amplitude of these inhomogeneous states is of order 0.1 $e$V (see Fig.~\ref{fig:Delta12} as a reference),
which is unphysically large as compared to the gaps observed experimentally. Furthermore, with such a
large superconducting gap, not only the electrons in the vicinity of the Fermi energy, but also those
far away from it, participate in the pairing process and contribute significantly to the condensation
energy. Therefore, the argument that the Fermi surface needs to match the gap structural factor\cite{Hu2012c}
does not apply here. As a consequence, it is justified that our focus is only on the case of small
exchanges $J$, in which the gap amplitudes, being of the order of 0.01 $e$V, are comparable to the experimental values.

\section{Discussion and Summary}
\label{sec:conclusion}

\begin{figure}
  \centering
  \includegraphics[width=8cm]{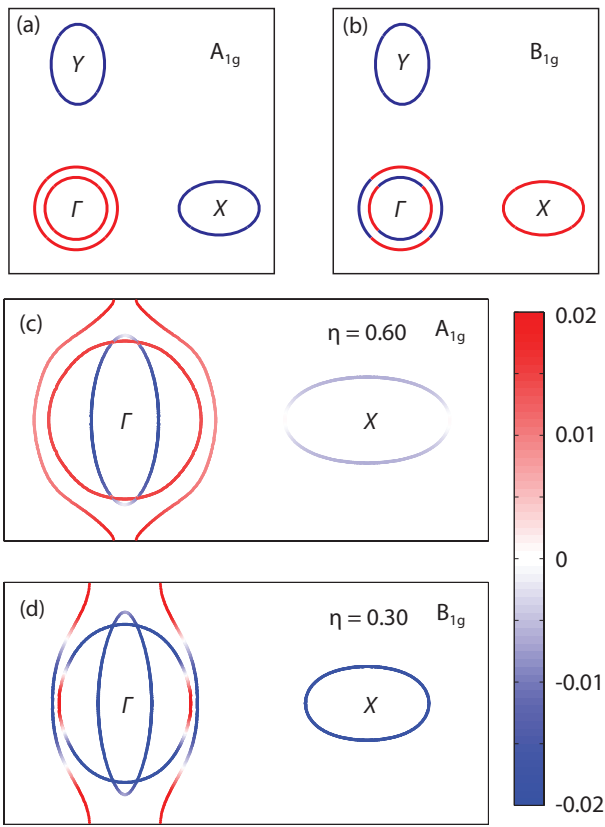}
  \caption{(Color online) (a, b) Illustration of the
superconducting order parameters $\Delta(k)$
on the Fermi surfaces of the two-dimensional Fe planes. Although (a) and (b) are
schematic, they are close to the Fermi surfaces for the model used here.
(a) $A_{1g}$ symmetry. $\Delta(k)$ is positive (red)
on the two hole pockets at $\Gamma$, and negative (blue) on the two electron pockets at $X$ and $Y$.
(b) $B_{1g}$ symmetry. $\Delta(k)$ changes sign under a $\pi/2$ rotation, with gap nodes on the hole pockets.
(c, d) The calculated superconducting order parameters $\Delta(k)$ on the Fermi surface of the coupled two-leg
iron ladders. (c) The $A_{1g}$ phase at $\eta = 0.6$.
(d) The $B_{1g}$ phase at $\eta=0.3$. The same exchange parameters
as in Fig.~\ref{fig:etaNNN1} have been used, i.e.~$J_1=0$ and $J_2=0.1D$,
with the critical coupling strength being $\eta_c \approx 0.41$.}
  \label{fig:fsop}
\end{figure}

In Sec.~\ref{sec:results}, it has been shown that the superconducting state of the two-leg Fe ladders
is dominated by the $B_{1g}$ phase, instead of the $A_{1g}$ phase found in the two-dimensional Fe planes.
But as it is well known, the superconducting order parameter of the $B_{1g}$ phase is required to change
sign under a $\pi/2$ rotation and gap nodes will always appear on the hole pockets at the zone center [Fig.~\ref{fig:fsop}(b)].
Consequently the $B_{1g}$ state usually has a higher energy than the nodeless $A_{1g}$ state [Fig.~\ref{fig:fsop}(a)].
So why does the $B_{1g}$ symmetry become the favorable pairing channel in the Fe-based ladders?

To understand this curious result, let us consider the effect introduced by an infinitesimal
change of $\eta$ from the isotropic limit of $\eta=1$. This perturbation hybridizes the states separated
by the momentum $(0,\pi)$ and effectively folds the Brillouin zone along the line $k_y = \pm \pi/2$. For
the general Fermi surface topology of the Fe-based superconductors [see Fig.~\ref{fig:fsop}(a) or (b)],
the electron pocket at $Y$ will overlap with the hole pockets at $\Gamma$ upon folding. As it can be seen
from Fig.~\ref{fig:fsop}(a), because the superconducting order parameters of the $A_{1g}$ state have
opposite signs on the hole and electron pockets, this hybridization generally reduces the gap amplitudes
and increases the energy of the system. By contrast, for the $B_{1g}$ state as shown in Fig.~\ref{fig:fsop}(b),
gap nodes already exist on the hole pockets. The hybridization between the hole and electron pockets will not
induce additional nodes, and thus the condensation energy of the system will remain the same up to the zeroth order. As a result,
as the Fe plane is decoupled into the two-leg ladders, at some critical value of $\eta=\eta_c$, the strength of the
hybridization may become strong enough so that the ground state will change from the $A_{1g}$ to the $B_{1g}$ pairing symmetry.

To support this argument, we plot in Figs.~\ref{fig:fsop}(c) and (d) the superconducting order parameters\cite{[{Note that
only the dominant intra-band superconducting order parameters are plotted. In general, both intra-band and inter-band pairings
are present after the transformation from the orbital to band basis. See }] Moreo2009a} on the Fermi surfaces of the coupled Fe-based
ladders in the folded Brillouin zone at two typical values of $\eta$. We have used the same exchange constants
($J_1=0$ and $J_2=0.1D$) as in Fig.~\ref{fig:etaNNN1}, where the symmetry change occurs at $\eta_c \approx 0.41$.
For $\eta = 0.6 > \eta_c$, the system retains the $A_{1g}$ symmetry of the isotropic planes [Fig.~\ref{fig:fsop}(c)].
But the gap amplitudes are significantly reduced due to the hybridization. For $\eta=0.3<\eta_c$, the Fermi surface pockets are more one-dimensional as compared to those at $\eta=0.6$. Eventually they will become straight lines along $k_y$ and the electron pocket at $X$ will disappear when $\eta$ approaches 0,
as shown in Fig.~\ref{fig:band}.
More importantly for our discussion, the dominant pairing
channel now has the $B_{1g}$ symmetry, with the gap nodes appearing on the hybridized hole pockets [Fig.~\ref{fig:fsop}(d)].
However, contrary to the common expectation from Fig.~\ref{fig:fsop}(b), the superconducting order
parameters have the same sign on the two hybridized electron pockets. This peculiar gap structure
simply manifests the fact that $A_{1g}$ and $B_{1g}$ are not rigorous representations for the
ladder system without the $D_{4h}$ symmetry so that different symmetry components are always mixed. 
In our model, it is the gap structure on the hybridized hole pockets that determines the leading pairing
symmetry of the whole system, which we identify as $A_{1g}$ for $\eta>\eta_c$ and $B_{1g}$ for $\eta<\eta_c$.

Furthermore, we can apply the same argument to the cuprates and understand why the $d$-wave pairing symmetry is favored in both Cu planes and ladders. For the Fermi surface topology of the cuprates, the largest gap amplitude appears at the anti-nodal regions, $(\pm \pi,0)$ and $(0, \pm \pi)$. Because these regions do not overlap with each other upon folding along $k_y = \pm \pi/2$, the hybridization does not cost significant energy despite of the sign change of the superconducting order parameters. Consequently, the dominant pairing symmetry remains $d$-wave as the Cu plane is decoupled into the two-leg ladders.

Such a change of the pairing symmetry induced by the hybridization effect has been
considered previously\cite{Mazin2011a,Khodas2012}
in the context of $\rm{KFe_2Se_2}$, where the hybridization occurs between the two electron pockets. In particular,
Ref.~\onlinecite{Khodas2012} shows that as the strength of the hybridization increases, the system goes from $d$-wave
to $s$-wave, with an intermediate $s+id$ state. This issue of time-reversal symmetry breaking is beyond the scope of the
current work because the superconducting order parameters are restricted to be real here. However, both phenomenological
and microscopic theoretical studies\cite{Lee2009c,Stanev2010,Maiti2013} have found that the appearance of the time-reversal
symmetry breaking phase, at least at the mean-field level, is very general for systems with competing pairing symmetries.
So it is possible that our model may
exhibit an intermediate $A_{1g}+iB_{1g}$ mean-field phase as the system dimension is reduced from planes to ladders.

In conclusion, we have studied the superconducting phase diagram of a three-orbital $t$-$J$ model defined
on two-leg ladders motivated by the recently discovered ladder iron selenides superconductors.
In contrast to the $A_{1g}$ state found in the two-dimensional Fe planes, the favorable pairing channel in the case of the
Fe-based ladders studied here has the $B_{1g}$ symmetry, which is characterized by the opposite signs of the superconducting
order parameters on the $d_{xz}$ and $d_{yz}$ orbitals. Furthermore, by investigating the dimensional crossover
from planes to ladders, we have found that the system undergoes a first-order transition between the two competing phases.
This change of the pairing symmetry may also occur continuously, with an intermediate $A_{1g} + iB_{1g}$ state that breaks the time-reversal symmetry.

\begin{acknowledgments}
This work was supported by the National Science Foundation Grant No.~DMR-1104386.
\end{acknowledgments}


\end{document}